\documentclass[fleqn,usenatbib]{mnras}

\usepackage{newtxtext,newtxmath}
\usepackage[T1]{fontenc}
\DeclareRobustCommand{\VAN}[3]{#2}
\let\VANthebibliography\thebibliography
\def\thebibliography{\DeclareRobustCommand{\VAN}[3]{##3}\VANthebibliography}

\usepackage{enumitem}
\usepackage{hyperref}
\usepackage{multirow}
\usepackage{soul}
\usepackage{threeparttable}
\usepackage{changepage}
\usepackage{url}
\usepackage{rotating}
\usepackage{graphicx}	
\usepackage{amsmath}	

\title[Cloud-9 \ion{H}{i} Structure]{The nature of Cloud-9: a compact core embedded in a diffuse envelope}

\author[R. Zhou et al.]{
Ruilei Zhou,$^{1,2}$\thanks{E-mail: zhourl@bao.ac.cn}
Ming Zhu,$^{1,2,3}$\thanks{E-mail: mz@nao.cas.cn}
Chuan-Peng Zhang,$^{1,3}$
and Jinlong Xu,$^{1,3}$
\\
$^{1}$State Key Laboratory of Radio Astronomy and Technology, National Astronomical Observatories, Chinese Academy of Sciences, Beijing 100101, China\\
$^{2}$University of Chinese Academy of Sciences, Beijing 100049, China\\
$^{3}$Guizhou Radio Astronomical Observatory, Guizhou University, Guiyang 550000, People's Republic of China
}

\date{Accepted XXX. Received YYY; in original form ZZZ}

\pubyear{2026}

\begin{document}
\label{firstpage}
\pagerange{\pageref{firstpage}--\pageref{lastpage}}
\maketitle

\begin{abstract}
We present new \ion{H}{i} observations of Cloud-9 with the Five-hundred-meter Aperture Spherical Telescope (FAST), measuring total flux of $0.35 \pm 0.03~\mathrm{Jy~km~s^{-1}}$. Combined with Very Large Array  (VLA) data, the \ion{H}{i} shows a two-component structure: a compact, kinematically quiescent core and an extended envelope. The outer component exhibits a coherent velocity gradient of $\sim25~\mathrm{km~s^{-1}}$ across $\sim7$~kpc. The position–velocity structure does not support rotation; instead, the gradient is consistent with ordered large-scale motion, induced by environmental interaction with M94.
The coexistence of a compact, kinematically quiescent core and an extended structured envelope is difficult to reconcile with a unbound or purely gaseous system. While the extended component is more naturally interpreted as environmentally shaped gas, the compact core suggests gravitational confinement beyond that provided by the observed baryonic content, without requiring strict long-term dynamical equilibrium. 
Deep optical limits further rule out a significant stellar component. Taken together, our results indicate that Cloud-9 is most naturally interpreted as a dark matter-dominated system undergoing environmental interaction. The observed morphology and kinematics are consistent with a stripped RELHIC scenario, in which a compact gas-rich core is embedded within an extended envelope shaped by ram-pressure interaction with the circumgalactic medium of M94.
\end{abstract}




\section{Introduction}
The $\Lambda$ cold dark matter ($\Lambda$CDM) paradigm predicts a steeply rising halo mass function toward low masses \citep{jenkins2001mass}. However, only a small fraction of these halos host observable galaxies. Numerical studies show that after reionization a characteristic mass threshold, $M_{\mathrm{crit}}(z)$, emerges from the balance between gravitational confinement, gas cooling, and heating by the ultraviolet background (UVB); only halos above this threshold can efficiently form stars \citep{hoeft2006dwarf,sawala2016chosen,benitez2020detailed}. At $z\sim0$, this threshold is $M_{\mathrm{crit}}\sim10^{9.7} M_{\odot}$ \citep{benitez2020detailed,nebrin2023starbursts}. As a result, a large population of low-mass halos is expected to remain dark, potentially retaining gas without forming stars.

Detecting gas-bearing, starless halos would provide direct evidence for dark matter structures at subgalactic scales. In this regime, the gas is expected to be pressure-supported within the halo potential and regulated by the UVB, producing relatively narrow \ion{H}{i} line widths and compact neutral cores embedded in more extended diffuse gas. Such configurations have been discussed in the context of reionization-limited \ion{H}{i} clouds (RELHICs), in which neutral gas is retained within low-mass dark matter halos after reionization \citep{benitez2017properties}. In the original simulations, RELHICs were modeled as isolated, pressure-supported gas clouds confined within low-mass dark matter halos. However, alternative interpretations, including tidal debris produced by galaxy interactions \citep{2013LNP...861..327D}, high-velocity clouds in galactic halos \citep{1997ApJ...488..216W}, or transient gaseous structures shaped by environmental effects \citep{putman2012galactic}, have also been proposed for starless \ion{H}{i} clouds.

Cloud-9, discovered by \citet{zhou2023fast} with FAST near M94, is a promising candidate for such a system. Its systemic velocity ($304~\mathrm{km~s^{-1}}$) is consistent with that of M94, and its projected separation is $\sim70$ kpc. The observed line width satisfies $W_{50}\lesssim20~\mathrm{km~s^{-1}}$, and, if located at the distance of M94, its \ion{H}{i} mass is $\sim10^6~M_{\odot}$ \citep{zhou2023fast,benitez2024examining}. Maintaining gravitational confinement at this scale requires a total mass significantly exceeding that of the detected gas. High-resolution VLA observations reveal a compact \ion{H}{i} core with no clear rotational signature and mildly distorted outer structure, suggesting environmental influence \citep{benitez2024examining}. Deep \textit{Hubble Space Telescope} imaging further rules out a stellar counterpart down to $\sim10^{3.5} M_{\odot}$ \citep{anand2025first}.

The RELHIC interpretation originates from simulations of isolated, low-mass halos in which gas is pressure-supported and regulated by the UV background \citep{benitez2017properties}. However, Cloud-9 appears more spatially extended than predicted for canonical RELHICs, as noted by \citet{2023ApJ...956....1B}, suggesting deviations from the isolated-halo assumption. Such deviations are expected if environmental effects are present, as also discussed by \citet{benitez2024examining}. While a RELHIC origin cannot be ruled out, distinguishing between an isolated and environmentally perturbed system requires further modelling beyond the original framework.

In this work, we present new FAST observations with improved spatial sampling and sensitivity. These data allow us to map the extended \ion{H}{i} distribution and velocity field of Cloud-9, providing new constraints on its dynamical state and the nature of its underlying gravitational potential. Section~2 describes the observations and data reduction, Section~3 presents the results, Section~4 discusses the origin of Cloud-9, and Section~5 summarizes our conclusions.

\section{Observations and Data Reduction}
Cloud-9 was observed with the FAST during two sessions on 2023 October 5 and October 22. FAST has a physical aperture of 500~m and an illuminated diameter of 300~m. The observations employed the 19-beam L-band receiver covering 1050--1450~MHz, together with the Spec (W) backend providing a bandwidth of 500~MHz divided into 65,536 spectral channels, corresponding to a velocity resolution of 1.67~km~s$^{-1}$. We adopted the multi-beam drift-scan mapping mode. The region was scanned along both the right ascension and declination directions. The receiver platform was rotated by $53\fdg4$ and $23\fdg4$ for the two scans, respectively, and the scan rate was set to $15\arcsec~\mathrm{s}^{-1}$ in both cases, resulting in more uniform sky coverage. To improve spatial sampling relative to previous FAST observations of Cloud-9, the scan spacing was reduced to 1\farcm17, significantly smaller than the 21\farcm6 used previously. At 1.4~GHz, the half-power beam width is $\sim$2\farcm9, and the pointing accuracy is approximately 12\arcsec.

Flux calibration was performed by injecting a 10~K noise diode for 1~s every 32~s, converting the data to antenna temperature units. Data reduction was carried out using the HIFAST pipeline \citep{jing2024hifast}, including baseline subtraction with an asymmetric weighted penalized least-squares algorithm. The calibrated spectra were gridded into a FITS data cube with a spatial pixel size of 1\arcmin. The final cube reaches an rms sensitivity of 5.66~mK per channel (0.37~mJy~beam$^{-1}$). For comparison, the rms sensitivity of the data cube of Cloud-9 presented by \citet{zhou2023fast} is 9.3~mK (0.6~mJy~beam$^{-1}$). Cloud-9 is also covered by the FASHI survey \citep{zhang2024fast, 2026arXiv260631539Z}, which has a spectral resolution of $\sim6.4~\mathrm{km~s^{-1}}$. Differences in spectral resolution may lead to variations in the data processing.

\section{Results}
\subsection{Extended \ion{H}{i} emission and flux recovery}
\begin{figure}
\begin{center}
	\includegraphics[width=0.4\textwidth]{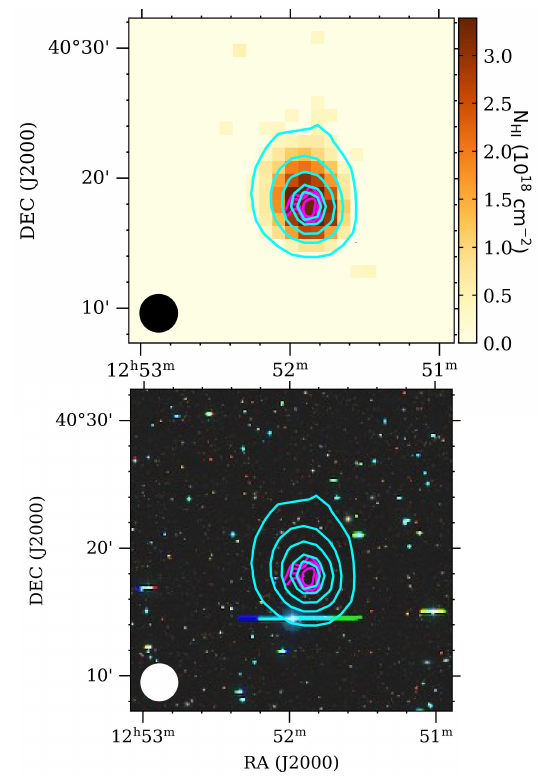}
    \caption{
    \textit{Top:} Integrated \ion{H}{i} column density map of the Cloud-9 region. 
    The cyan contours show the FAST data from this work, integrated over 
    $265.40$--$332.03~\mathrm{km~s^{-1}}$. 
    The contours start at $3\sigma$ level ($2.97 \times 10^{17}~\mathrm{cm^{-2}}$) 
    and increase in steps of $10\sigma$($9.99 \times 10^{17}~\mathrm{cm^{-2}}$). 
    The magenta contours correspond to the VLA data from \citet{benitez2024examining}, integrated over 
    $293.00$--$314.00~\mathrm{km~s^{-1}}$, with a starting level of 
    $3\sigma$ ($5.45 \times 10^{18}~\mathrm{cm^{-2}}$); the contours increase inward by factors of $\sqrt{2}$. 
    \textit{Bottom:} The same FAST and VLA contours as the top panel overlaid on the DESI-LS optical image. 
    The circle shown in the lower-left corner of each panel represents the FAST beam size.
     }
\label{fig:moment0}
\end{center}
\end{figure}

Figure~\ref{fig:moment0} presents the integrated \ion{H}{i} column density map of Cloud-9 derived from the FAST data cube over the velocity range $265.40$--$332.03~\mathrm{km,s^{-1}}$. For comparison, the VLA contour map from \citet{benitez2024examining} is also overlaid on it. The FAST emission is detected down to a $3\sigma$ column density level of $2.97 \times 10^{17}~\mathrm{cm^{-2}}$, nearly an order of magnitude lower than that of the VLA observations ($3\sigma = 5.45 \times 10^{18}~\mathrm{cm^{-2}}$ ), and improved relative to previous FAST measurements \citep{zhou2023fast}. 

At this sensitivity, the \ion{H}{i} emission extends beyond the region traced by the VLA, revealing diffuse structure that is not recovered by the interferometric data. In particular, additional low–surface-brightness emission is detected toward the northern side of Cloud-9, indicating that the gas distribution is more extended than previously inferred. This extended morphology is qualitatively consistent with the interpretation proposed by \citet{benitez2024examining}, who suggested that discrepancies between single-dish and interferometric measurements may arise from diffuse \ion{H}{i} emission that is not recovered by the VLA.

\begin{figure*}
\begin{center}
	\includegraphics[width=0.6\textwidth]{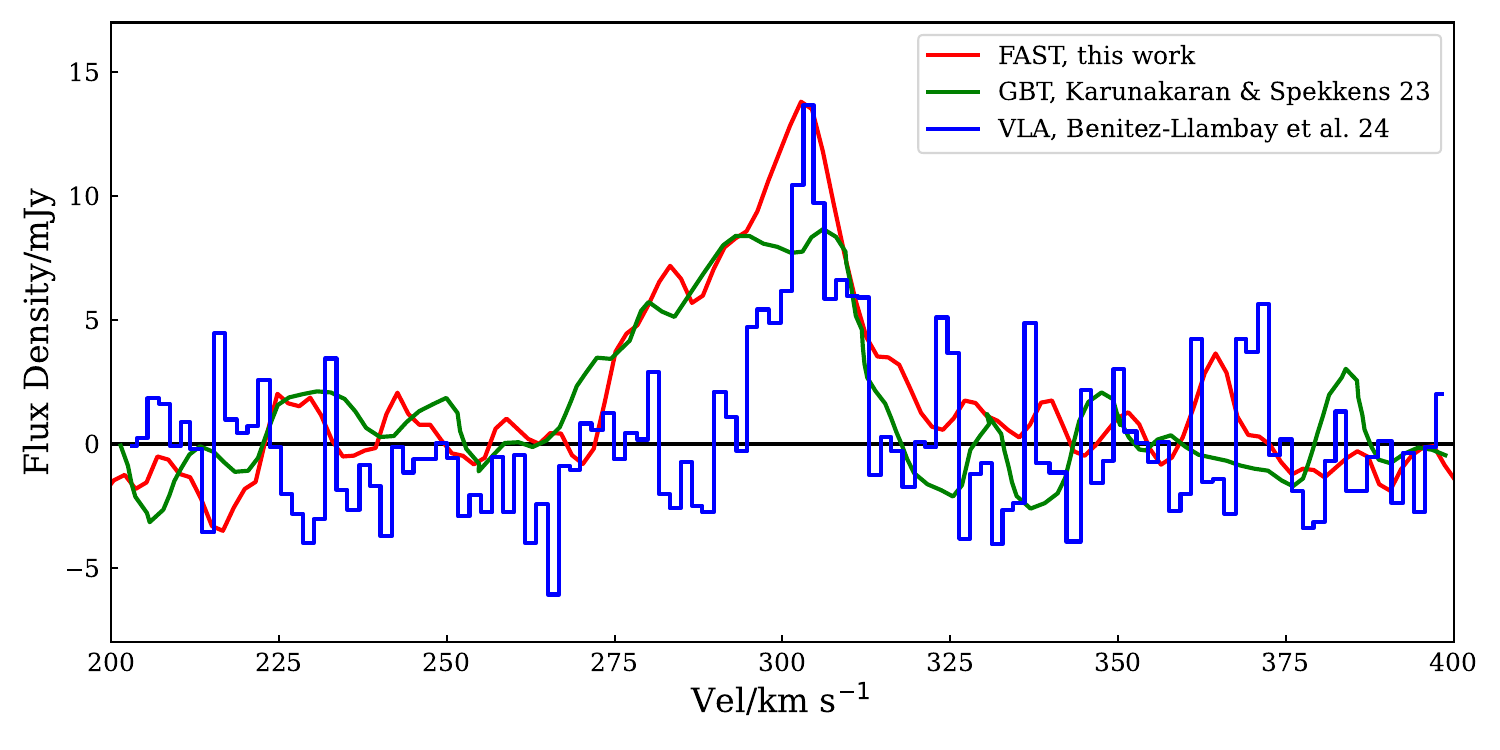}
    \caption{
    \ion{H}{i} spectral profiles of Cloud-9. The red solid line shows the \ion{H}{i} spectrum obtained with the FAST in this work. The blue solid line corresponds to the VLA-D spectrum presented by \citet{benitez2024examining}. The green solid line corresponds to \ion{H}{i} spectrum obtained with the GBT in \citet{karunakaran2024bigger}.
    }
\label{fig:flux}
\end{center}
\end{figure*}

The integrated \ion{H}{i} spectrum of Cloud-9 is shown in Figure~\ref{fig:flux}. The overall profile shape and centroid velocity are consistent with previous single-dish measurements, with a systemic velocity of $\sim 300~\mathrm{km\,s^{-1}}$. We measure a total flux of $0.35 \pm 0.03~\mathrm{Jy\,km\,s^{-1}}$, corresponding to an \ion{H}{i} mass of $(1.79 \pm 0.15) \times 10^{6}~M_\odot$ assuming a distance of 4.66~Mpc. 

This flux is significantly higher than the value derived from VLA observations ($0.16 \pm 0.02~\mathrm{Jy\,km\,s^{-1}}$; \citealt{benitez2024examining}). Given the consistency of the spectral profile with previous single-dish measurements, the excess flux recovered by FAST is unlikely to arise from calibration differences. Instead, it reflects the presence of extended, low–surface-brightness \ion{H}{i} emission that is missed by the interferometric observations.

Together, these results confirm that a substantial fraction of the \ion{H}{i} emission in Cloud-9 resides in a diffuse component that is not captured by the VLA, supporting earlier suggestions based on discrepancies between single-dish and interferometric measurements and highlighting the importance of deep single-dish observations for recovering the full gas content of the system.

\subsection{Large-scale velocity gradient}
\begin{figure*}
\begin{center}
	\includegraphics[width=0.7\textwidth]{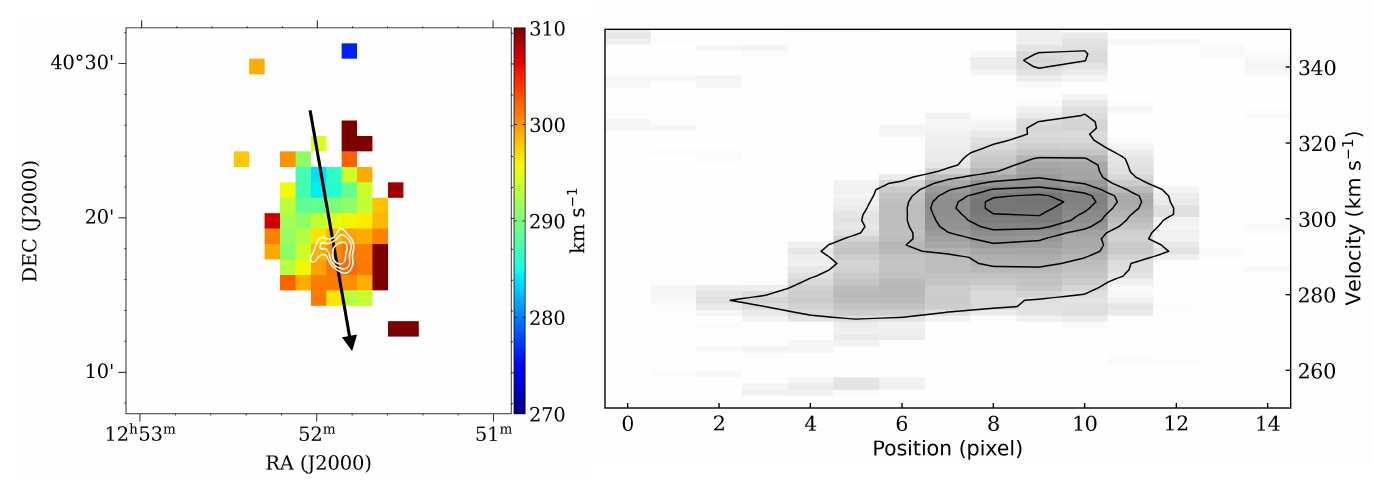}
    \caption{
    Position–velocity structure of Cloud-9. Left panel: intensity-weighted mean velocity (moment 1) map of Cloud-9 derived from the FAST data cube. The black arrow indicates the orientation and spatial extent of the slice used to extract the position–velocity (PV) diagram shown in the right panel. The half-power beam width (HPBW) of FAST is illustrated in the lower-left corner. White contours show the VLA \ion{H}{i} column density distribution for comparison. Right panel: PV diagram extracted along the direction marked in the left panel. Black contours start at 3$\sigma$ (1.11~mJy/beam) and increase in steps of 4$\sigma$.
    }
\label{fig:pv}
\end{center}
\end{figure*}

The \ion{H}{i} velocity field derived from the FAST data (left panel of Figure~\ref{fig:pv}) reveals a coherent gradient across Cloud-9, with velocities decreasing from the southwest to the northeast. This moment-1 map was generated from the spectral cube using a standard spectral cube algorithm\footnote{\url{https://spectral-cube.readthedocs.io/en/latest/moments.html\#moment-maps}} and smoothed with a $3\times3$ pixel Gaussian kernel. To ensure reliable velocity measurements, only regions with column densities above the $3\sigma$ level are shown. The corresponding position--velocity (PV) diagram (right panel of Figure~\ref{fig:pv}) shows a continuous velocity variation of $\sim 25~{\rm km~s^{-1}}$ over a projected scale of $\sim 7$~kpc.

The observed kinematic structure is inconsistent with ordered rotation. A rotating system is expected to exhibit a symmetric velocity field about a well-defined dynamical center and a characteristic double-sided structure in the PV diagram \citep[e.g.,][]{oh2015high,de2008high}. In contrast, the gradient in Cloud-9 is approximately linear, lacks a clear dynamical center, and does not show a symmetric velocity pattern or evidence of a flattened rotation curve. 

Beam smearing cannot produce such a large-scale coherent gradient, as it acts to smooth existing velocity structure rather than generate it. Furthermore, the compact \ion{H}{i} core resolved by VLA observations shows no measurable internal velocity gradient \citep{benitez2024examining}. 

These results indicate that the velocity field is dominated by large-scale ordered motion rather than rotational support, and that the kinematics of the extended gas differ from those of the central core.

\subsection{A two-component \ion{H}{i} structure}
VLA observations resolve only a compact and dynamically quiescent core, which shows no significant internal velocity gradient \citep{benitez2024examining}. In contrast, FAST data reveal a more diffuse and spatially extended \ion{H}{i} envelope, exhibiting a coherent velocity gradient on this larger scale. This naturally points to a layered structure: a dynamically quiescent, compact core embedded within an outer component with organized kinematics. The orientation of the large-scale velocity gradient does not follow the morphology of the compact core, suggesting that the two components are kinematically distinct. This comparison is illustrated in Figure~\ref{fig:pv}, where the compact component traced by the VLA emission is overlaid on the FAST velocity field. The total \ion{H}{i} flux measured by FAST is significantly higher than that from VLA, which is attributed to spatial filtering effects inherent to interferometric observations \citep{wang2024feasts}, supporting the interpretation that the double-layered structure is an intrinsic property of Cloud-9.

\begin{figure*}
\begin{center}
	\includegraphics[width=0.6\textwidth]{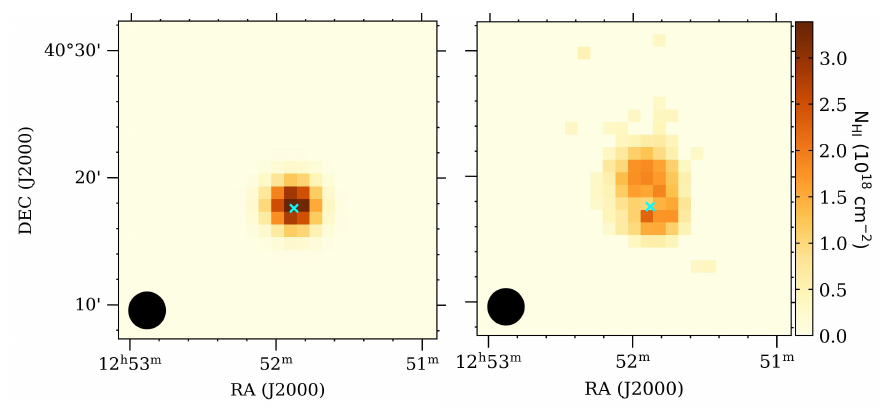}
    \caption{
    Left: VLA data convolved to the angular resolution of FAST. Right: residual \ion{H}{i} emission obtained by subtracting the convolved VLA data from the FAST data (Figure~\ref{fig:moment0}), highlighting the extended, low-surface-brightness component not recovered by the interferometric observations. The cyan crosses in each panel mark the center of Cloud-9.
    }
\label{fig:FAST_VLA}
\end{center}
\end{figure*}

To clearly reveal the spatial separation of these components, we convolved the VLA data cube to the FAST resolution and subtracted it from the FAST data. As shown in Figure~\ref{fig:FAST_VLA}, the residual emission is predominantly located outside the central region, with a clear excess toward the northern side of Cloud-9. This confirms that the compact core and extended emission are spatially distinct. Note that the VLA observations also covered the northern part of this cloud. The non-detection by the VLA confirmed that this region contains pure diffuse gas without any detectable dense clumps. 

\begin{figure*}
\begin{center}
	\includegraphics[width=0.6\textwidth]{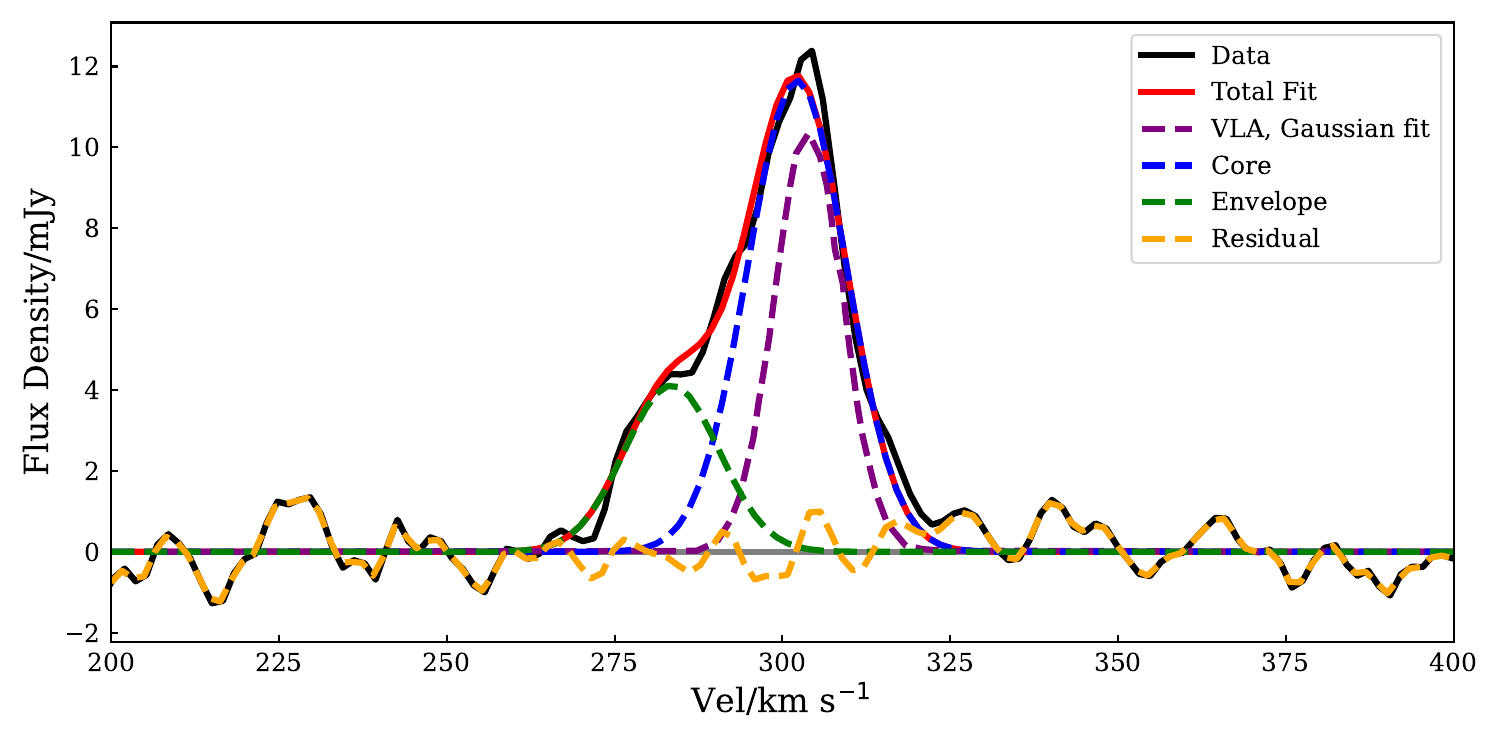}
    \caption{
    The black solid line shows the \ion{H}{i} profile obtained by integrating over the specific region and velocity range described in Section 3.3. The red solid line represents the total Gaussian fit to this \ion{H}{i} profile. The blue, green, and orange dashed lines represent the fitted core, envelope, and residual components, respectively. And the purple dashed line shows the Gaussian fit to the VLA \ion{H}{i} profile from \citet{benitez2024examining}.
    }
\label{fig:profile_fit}
\end{center}
\end{figure*}

The two components are further distinguished in the spectral domain, as shown in Figure~\ref{fig:profile_fit}. For the FAST data cube, we extracted the \ion{H}{i} profile by integrating over the area corresponding to the VLA emission convolved to the FAST resolution (Figure~\ref{fig:FAST_VLA}, left panel).
Rather than performing a simultaneous double-Gaussian fit, we decompose the integrated \ion{H}{i} profile into two components. The central, narrow component is modeled with a single Gaussian, representing the compact core, while the residual emission is attributed to the extended envelope and characterized separately.
The Gaussian fit to the core yields a centroid velocity consistent with the VLA measurements and a line width of $W_{50} = 17.34 \pm 0.83~\mathrm{km,s^{-1}}$. The remaining flux, not accounted for by the core component, corresponds to the broader, lower-amplitude emission recovered by FAST. The derived parameters are summarized in Table~\ref{table:fit}.

The kinematic contrast between the two components is particularly striking. The inner core shows no measurable velocity gradient, whereas the extended component exhibits a coherent large-scale gradient whose orientation differs from the morphology of the compact core. Combined with the spatial separation revealed by the residual map, this suggests that the two components represent distinct dynamical structures.

Taken together, these results demonstrate that Cloud-9 consists of a compact, dynamically quiescent core embedded within a more extended diffuse envelope with distinct spatial and kinematic properties.

\begin{table}
\centering
\caption{Parameters of the core and envelope components}
\label{table:fit}
\begin{tabular}{c|cccc}
\hline
 & $S_{\nu}$ & $M_{\mathrm{HI}}$ & $v_{\mathrm{helio}}$ & $W_{50}$ \\
 & $\mathrm{Jy\,km\,s^{-1}}$ & $\times 10^{6}~M_\odot$ & $\mathrm{km\,s^{-1}}$ & $\mathrm{km\,s^{-1}}$ \\
\hline
Core     & $0.22 \pm 0.01$ & $1.13 \pm 0.05$ & $302.58 \pm 0.45$ & $17.34 \pm 0.83$ \\
Envelope & $0.13 \pm 0.03$ & $0.67 \pm 0.15$ & --- & --- \\
\hline
\end{tabular}

\medskip
\noindent\begin{minipage}{\linewidth}
\textbf{Note.} Fluxes are measured from the integrated \ion{H}{i} profiles. The \ion{H}{i} masses are calculated as $M_{\mathrm{HI}} = 2.36 \times 10^{5} D^{2} \int S\,dv$, where $D$ is the distance in Mpc. The centroid velocity is defined as the flux-weighted velocity, and $W_{50}$ denotes the full width at half maximum of the profile. The envelope component is obtained by subtracting the fitted core from the total \ion{H}{i} profile.
\end{minipage}
\end{table}

\section{Implications for the nature of Cloud-9}
Cloud-9 exhibits a two-component \ion{H}{i} structure: a compact, dynamically quiescent core embedded in a diffuse, extended envelope with a coherent velocity gradient. Such a configuration is inconsistent with typical isolated HVCs, which display broad, turbulent motions rather than stable, ordered gradients \citep{wakker1997high}. Similarly, double-layered purely gaseous clouds in the Local Group or Galactic supershells form via instabilities, with dense clumps showing axis-aligned or turbulent gradients and extended components exhibiting radial expansion rather than coherent linear motion \citep{heiles1984hi,putman2003magellanic}.

The inner core of Cloud-9 exhibits a velocity dispersion of 
$\sim8$--$9~\mathrm{km~s^{-1}}$ and an \ion{H}{i} mass of 
$\sim10^6~M_\odot$. Such a line width implies that the core is 
dynamically quiescent, with a relatively low level of turbulent motions, and is therefore unlikely to be a freely expanding, unbound gaseous structure. However, given the ongoing ram-pressure stripping by M94's  circumgalactic medium, we do not assume that the core is in strict or  long-term dynamical equilibrium. 
In contrast, the extended envelope shows a coherent velocity gradient of $\sim25~\mathrm{km~s^{-1}}$ across $\sim7$ kpc, suggesting that the envelope is actively being stripped by M94’s circumgalactic medium. Hydrodynamical simulations of HVCs of comparable mass indicate that a pure HI cloud core would be rapidly disrupted by ram pressure, instabilities, and ionization unless confined by a gravitational potential \citep{nichols2009smith,nichols2014smith}.
The persistence of a compact, dynamically quiescent core within this stripping environment therefore suggests that the inner gas remains at least temporarily confined by an underlying gravitational potential, most naturally associated with dark matter, even though the core itself may still evolve as the outer gas layers are being stripped away.

Timescale arguments reinforce this conclusion. Traveling $\sim40$–$50$ kpc from M94’s outer disk at typical halo velocities ($v\sim100$–$200~\mathrm{km~s^{-1}}$) implies $t_{\rm travel} \sim 0.2$–0.5 Gyr. Simulations of tidal debris or HVCs in similar environments predict disruption timescales of only $\sim0.1$ Gyr \citep{putman2012galactic,heitsch2009fate}. Without a confining potential, an external HVC or tidal origin cloud would be rapidly ionized and dispersed \citep{gnedin2000effect,benitez2017properties}.
We note that tidal debris and HVCs are physically distinct phenomena: HVCs are typically defined as gas clouds with velocities offset from the systemic velocity of the host galaxy by $\gtrsim 90~\mathrm{km~s^{-1}}$, whereas tidal debris originates from gravitational interactions and generally follows the kinematics of the parent system. Nevertheless, both scenarios involving unbound gas clouds require survival over comparable timescales in the circumgalactic medium.  
Alternatively, if Cloud-9 originated as an external HVC accreting onto the M94 system, it would still need to remain neutral while traveling through the circumgalactic environment for several hundred Myr. In the absence of a confining gravitational potential, such a cloud would be highly susceptible to ionization by the extragalactic UV background and to hydrodynamical disruption \citep[e.g.,][]{gnedin2000effect,benitez2017properties}.  

Within this framework, 
assuming the gas envelope was growing at a speed similar to the velocity gradient($\sim25~\mathrm{km~s^{-1}}$), the spatial span ($\sim$  7 kpc) yields a kinematic evolutionary age of $\sim$ 270 Myr, consistent with the independently derived orbital travel
time (0.2–0.5 Gyr).
This provides quantitative evidence that
the envelope is a transient tail growing over the duration of Cloud-9’s infall, consistent with
a “stripped RELHIC” interpretation. 

The baryonic content is tightly constrained. Deep \textit{HST} imaging rules out a stellar counterpart of $10^{3.5}~M_\odot$ at high confidence \citep{anand2025first}, and diffuse \ion{H}{i} at the observed column densities is not expected to be self-gravitating \citep[e.g.,][]{elmegreen1997pressure,putman2012galactic}. Together, these constraints indicate that baryons alone cannot provide the required gravitational support.

Taken together, Cloud-9 is most naturally interpreted as a compact, gravitationally bound core embedded in an extended envelope shaped by environmental processes, consistent with a stripped RELHIC-like system undergoing ram-pressure interaction in the circumgalactic medium of M94. This scenario is in line with the interpretation of \citet{benitez2024examining}, who proposed that Cloud-9 is a RELHIC-like system whose asymmetric morphology is driven by ram-pressure stripping.

While the extended envelope exhibits large-scale ordered kinematics, we emphasize that it does not trace the gravitational potential of the system. Instead, it reflects transient gas dynamics associated with the ongoing interaction between Cloud-9 and its environment, whereas the compact core provides the primary evidence for a dark matter-dominated potential.

\section{Conclusion}
\begin{enumerate}
\item \textbf{Two-component structure.}  
Combining VLA observations, FAST reveals Cloud-9 as a compact, kinematically quiescent core embedded within a more extended \ion{H}{i} envelope. The outer component exhibits a coherent large-scale velocity variation of $\sim 25~{\rm km~s^{-1}}$ across $\sim 7$~kpc, with an orientation misaligned with the inner core.

\item \textbf{Pure gas-cloud scenario disfavored.}
The compact, dynamically quiescent core and extended, kinematically disturbed envelope are difficult to reconcile with a completely unbound or purely gaseous system. Survival in the M94 environment further disfavors a pure gas-cloud origin for Cloud-9, although we do not assume that the core is in strict or long-term dynamical equilibrium.

\item \textbf{Dark matter halo and stripped RELHIC interpretation.}
The compact core is difficult to explain with the observed baryonic content alone and suggests gravitational confinement on the observed scales, while the extended envelope is consistent with gas shaped by environmental interaction.
Together, these properties support Cloud-9 being a stripped RELHIC-like dark matter halo.
\end{enumerate}

In summary, our FAST observations strongly disfavor purely gaseous or stellar origins for Cloud-9 and instead indicate that it is a dark matter-dominated system. The detection of an extended \ion{H}{i} envelope and non-equilibrium kinematics, likely shaped by interaction with M94, supports a scenario in which Cloud-9 is a stripped RELHIC object undergoing ram-pressure interaction within the circumgalactic medium. Future higher-resolution observations and numerical simulations will further constrain the dynamical structure and environmental evolution of this system.

\section*{Acknowledgements}

We thank Dr. Alejandro Benitez-Llambay for providing the VLA data. We also thank the anonymous referee for constructive comments and suggestions. This work is supported by the National SKA Program of China (2025SKA0150100), the National Natural Science Foundation of China (12373001), and the Guizhou Provincial Science and Technology Projects (QKHFQ[2023]003, QKHPTRCZDSYS[2023]003, QKHFQ[2024]001-1). FAST is a Chinese national mega-science facility operated by the National Astronomical Observatories of the Chinese Academy of Sciences (NAOC).

\section*{Data Availability}
The raw data used in the article will be published on the FAST website: \href{https://fast.bao.ac.cn}{https://fast.bao.ac.cn}.
The PID is N2023$\_$1.
Please contact the author (zhourl@bao.ac.cn) for processed data.


\bibliographystyle{mnras}
\bibliography{ref} 

\bsp	
\label{lastpage}
\end{document}